\documentclass[aps,twocolumn,nofootinbib]{revtex4}
\usepackage{bm,epsf}
\newcommand{\kB}{k_{\rm B}}
\newcommand{\Ass}{\mbox{\sf A}}

\newcommand{\ave}[1]{\left\langle#1\right\rangle_\rho}
\newcommand{\Dbra}[1]{\left\langle#1\right|}
\newcommand{\Dket}[1]{\left|#1\right\rangle}
\newcommand{\cancor}[2]{\left\langle#1;#2\right\rangle_\rho}
\newcommand{\Qcommu}[2]{[#1,#2]}
\newcommand{\QPoiss}[2]{(#1,#2)}
\newcommand{\Qantico}[2]{\{#1,#2\}}
\newcommand{\CPoiss}[2]{\bm{\{}#1,#2\bm{\}}}
\newcommand{\Cdissip}[2]{\mbox{$\bm{[\hspace{-0.28em}[}$}#1,#2\mbox{$\bm{]\hspace{-0.28em}]}$}}

\begin{document}
\bibliographystyle{apsrev}

\title{The geometry and thermodynamics of dissipative quantum systems}

\author{Hans Christian \"Ottinger}
\email[]{hco@mat.ethz.ch}
\homepage[]{http://www.polyphys.mat.ethz.ch/}
\affiliation{ETH Z\"urich, Department of Materials, Polymer Physics, HCI H 543,
CH-8093 Z\"urich, Switzerland}

\date{\today}

\begin{abstract}
Dirac's method of classical analogy is employed to incorporate quantum degrees of freedom into modern nonequilibrium thermodynamics. The proposed formulation of dissipative quantum mechanics builds entirely upon the geometric structures implied by commutators and canonical correlations. A lucid formulation of a nonlinear quantum master equation follows from the thermodynamic structure. Complex classical environments with internal structure can be handled readily.
\end{abstract}



\maketitle


\emph{Introduction.}---In the context of solid state-based quantum information processing, external control in quantum optics, quantum transport through meso- and nanoscale structures, quantum tunneling in macroscopic systems, quantum Brownian motors, as well as biological reactions and protein folding kinetics, for example, quantum dissipation is an important subject. We here propose a general thermodynamic framework for dissipative quantum systems in contact with classical equilibrium and nonequilibrium environments. The formulation is guided by classical nonequilibrium thermodynamics, based entirely on geometric concepts and, in principle, applicable to arbitrarily complex quantum subsystems and classical environments. Important is the weak coupling of a quantum system to its environment because otherwise thermodynamic concepts are clearly not applicable.

The proper arena for quantum mechanics is provided by Hilbert spaces, that is, complex vector spaces equipped with inner products \cite{Dirac,Schiff}. In the infinite dimensional case, a Hilbert space ${\cal H}$ needs to be sufficiently rich to tell a nice story with all its punchlines (completeness: every Cauchy sequence has a limit in the space ${\cal H}$) and sufficiently small to play with the tools of measure theory (separability: there exists a countable subset that is dense in ${\cal H}$). Observables are densely defined self-adjoint linear operators on a Hilbert space ${\cal H}$. As a consequence of self-adjointness, observables are real. The space $\Ass({\cal H})$ of quantum observables hence is a real vector space. With the quantum Poisson bracket implied by the commutator $\Qcommu{\,}{\,}$ as a binary operation, $\Ass({\cal H})$ becomes a Lie algebra.

For nonequilibrium systems, we are interested in the evolution of averages of a sufficiently rich set of observables. It is not immediately obvious what ``sufficiently rich'' means for the characterization of the state of a quantum system. We here focus on the evolution of the density matrix $\rho$, also known as the statistical operator, which can be expressed in terms of an orthonormal basis $\Dket{n}$ of the Hilbert space ${\cal H}$ and the probabilities $p_n$, namely, $\rho = \sum_n p_n \Dket{n} \Dbra{n}$. The density matrix characterizes the state of our quantum subsystem and its evolution determines the evolution of the averages of all observables. This point of view corresponds to the Schr\"odinger picture which we use throughout this letter.

The corresponding arena for classical systems is given by functions or functionals on a Poisson manifold as observables. Classical nonequilibrium thermodynamics provides evolution equations for the values of observables in terms of Poisson and dissipative brackets as the appropriate geometric structures \cite{BerisEdwards,hco99,hco100,hcobet}. In practice, one usually follows the evolution of a point $x$ in a Poisson manifold. The evaluation of observables as functions or functionals of $x$ is the classical counterpart of the averaging of quantum observables with a density matrix $\rho$.


\emph{The classical analogy.}---We here explore the possible mathematical setting for the formulation of nonequilibrium systems with quantum degrees of freedom. The treatment of such systems should be inspired by the nonequilibrium thermodynamics of purely classical systems. We therefore need formulations that emphasize the analogy between classical and quantum systems. This classical analogy was a major concern of P.A.M.~Dirac who, in his pioneering textbook on \emph{The Principles of Quantum Mechanics} \cite{Dirac}, devotes an entire chapter to the relationship between Poisson brackets and commutators which, in Chapter~IV of the book \cite{Dirac}, are referred to as the quantum conditions. According to Dirac, ``We should thus expect to find that important concepts in classical mechanics correspond to important concepts in quantum mechanics, and, from an understanding of the general nature of the analogy between classical and quantum mechanics, we may hope to get laws and theorems in quantum mechanics appearing as simple generalizations of well-known results in classical mechanics'' (p.~84 of \cite{Dirac}). Following Dirac, we here rely on his method of classical analogy.

It is important to realize that, by means of the classical analogy and simple geometric concepts, we wish to formulate nonequilibrium thermodynamics, not to derive it. Rigorous derivations, such as ergodic theorems, are almost impossible to achieve even in classical equilibrium thermodynamics. Finding the proper formulation of nonequilibrium thermodynamics is certainly much simpler than deriving it, but no less useful. Once the proper thermodynamic setting is available, models for specific applications can be formulated readily. As in equilibrium thermodynamics, closed systems, here consisting of quantum subsystems and their classical environments, are most suited for fundamental developments and conceptual clarity.


\emph{Reversible and irreversible structures.}---A general formulation of nonequilibrium thermodynamics based on geometric concepts has been developed during the last decades \cite{BerisEdwards,hco99,hco100,hcobet}. A key idea is the need for separate geometric structures for generating reversible and irreversible dynamics.

The proper structure underlying reversible dynamics is given known to be given by the quantum Poisson bracket of two observables $A$ and $B$ in terms of their commutator and Planck's constant $\hbar$,
\begin{equation}\label{QPoiss}
    \QPoiss{A}{B} = \frac{1}{i \hbar} \, \Qcommu{A}{B} ,
\end{equation}
carefully motivated and introduced by Dirac (see p.~87 of \cite{Dirac}), and here written in the notation of the textbook \cite{KuboetalII}. The rate of reversible change of the average $\ave{A} = {\rm tr}(A \rho)$ of a quantum observable $A$ is given by the average $\ave{\QPoiss{A}{H}}$, where the observable $H$ is the Hamiltonian of the quantum system. The Poisson bracket has been recognized by Dirac as the common geometric structure underlying reversible dynamics in both classical and quantum systems. The entire reversible structure follows from the noncommuting nature of quantum observables. Moreover, according to Heisenberg's uncertainty principle, the averaged quantum Poisson bracket $\ave{\QPoiss{A}{B}}$ occurs in the lower bound for the product of the uncertainties in two variables $A$ and $B$; it certainly provides a most fundamental structure.

In the same spirit, we next propose a simple geometric structure behind all irreversible dynamics. The essential structural element is $\cancor{\QPoiss{A}{Q}}{\QPoiss{B}{Q}}$, where $Q$ is a suitable observable that provides the dissipative coupling of the quantum subsystem to its environment. The bilinear pairing $\cancor{\,}{\,}$ in this structural element is the canonical correlation (see Eq.~(4.1.12) of \cite{KuboetalII})
\begin{equation}\label{cancor}
    \cancor{A}{B} = \int_0^1 {\rm tr}
    \big( \rho^\lambda A \, \rho^{1-\lambda} B \big) \, d\lambda
    = {\rm tr} \big( A_\rho B \big) ,
\end{equation}
where
\begin{equation}\label{Atildef}
    A_\rho = \int_0^1  \rho^\lambda A \, \rho^{1-\lambda} \, d\lambda ,
\end{equation}
is basically the product of $A$ and $\rho$ but, as we need to compromise between writing $\rho$ to the left or the right of $A$, we place it as a subscript. The canonical correlation is symmetric, $\cancor{A}{B} = \cancor{B}{A}$, and positive, $\cancor{A}{B} \geq 0$. Additional convexity properties can be inferred from Lieb's theorem (see, for example, Eq.~(2.120) of \cite{BreuerPetru}). Finally, the formula $\cancor{A}{1} = \ave{A}$ shows that canonical correlation may be considered as a pairing of observables and measures (density operators modifying the statistical operator $\rho$ in the spirit of Radon-Nikodym derivatives), so that the dual spaces of observables and measures have been brought on an entirely equal footing.

In the very same way that the reversible structure $\ave{\QPoiss{A}{B}}$ assists the energy to generate reversible evolution, irreversible evolution should be generated by entropy by means of the irreversible geometric structure $\cancor{\QPoiss{A}{Q}}{\QPoiss{B}{Q}}$. Assuming von Neumann's logarithmic form of the entropy, this possibility relies on the following lemma,
\begin{equation}\label{lnLemma}
    \QPoiss{\ln\rho}{A_\rho} = \QPoiss{\rho}{A} .
\end{equation}
To prove this result, we can look at arbitrary matrix elements formed with the eigenstates of the density matrix,
\begin{eqnarray}
    ( \ln p_n - \ln p_m ) \Dbra{n} A_\rho \Dket{m} =
    p_m \ln \frac{p_n}{p_m} \int_0^1 \left( \frac{p_n}{p_m} \right)^\lambda d\lambda
    && \nonumber \\
    \times \Dbra{n} A \Dket{m} = ( p_n - p_m ) \Dbra{n} A \Dket{m} . \qquad \qquad &&
\label{lnLemmaproof}
\end{eqnarray}
By means of the lemma (\ref{lnLemma}), we obtain the equally useful and elegant formula
\begin{equation}\label{doubcomid}
    - \cancor{\QPoiss{A}{Q}}{\QPoiss{\ln\rho}{Q}} = \ave{\QPoiss{Q}{\QPoiss{Q}{A}}} .
\end{equation}


\emph{Evolution equations.}---To obtain the total energy and entropy of a closed system we need to know the corresponding pairs $\tilde{A} = (A, A_{\rm e})$ of quantum and classical observables, so that one can write
\begin{equation}\label{Agenave}
    \bar{A} = \ave{A} + A_{{\rm e},x} .
\end{equation}
The subscript $x$ indicates that an observable is evaluated in the classical state $x$. Had we used probability measures on the state space as independent classical variables (see Sec.~6.3 of \cite{hcobet}), the appearance of quantum and classical contributions would be even more similar, but this level of detail is not necessary for our purposes and most applications. For the energy $\tilde{H}$, $A$ is the Hamiltonian $H$ of the quantum subsystem and $A_{\rm e}$ is the energy $H_{\rm e}$ (or $E_{\rm e}$) of the environment. For the entropy $\tilde{S}$, we choose the operator $A$ as $S=-\kB\ln\rho$, where $\kB$ is Boltzmann's constant, and $A_{\rm e}$ is the entropy $S_{\rm e}$ of the environment. We can then formulate the evolution of the average of any joint observable of the quantum system and its environment in terms of the generators $\tilde{H}$ and $\tilde{S}$,
\begin{equation}\label{GENERIC}
    \frac{d\bar{A}}{dt} = {\cal P}(\tilde{A},\tilde{H})
    + {\cal D}(\tilde{A},\tilde{S}) ,
\end{equation}
where both the reversible Poisson contribution ${\cal P}$ and the irreversible dissipative contribution ${\cal D}$ consist of classical and quantum contributions. This formulation is the most natural generalization of the GENERIC framework \cite{BerisEdwards,hco99,hco100,hcobet} of classical nonequilibrium thermodynamics (``general equation for the nonequilibrium reversible-irreversible coupling''). We have
\begin{equation}\label{calPdef}
    {\cal P}(\tilde{A},\tilde{B}) = \CPoiss{A_{\rm e}}{B_{\rm e}}_x + \ave{\QPoiss{A}{B}}  ,
\end{equation}
where $\CPoiss{\,}{\,}$ is the classical Poisson bracket. Unfortunately, square and curly brackets have been used to distinguish between commutators and anticommutators in quantum mechanics, between quantum and classical Poisson brackets, and between Poisson and dissipative brackets in classical nonequilibrium thermodynamics. The same symbols occur with different meanings. Although the distinction between classical Poisson brackets and quantum anticommutators (both expressed as curly brackets) as well as between classical dissipative brackets and quantum commutators (both expressed as square brackets) should always be clear from the context, we here try to facilitate the distinction by using boldface brackets for the classical objects.

For the dissipative contribution to evolution we employ the following geometric structure,
\begin{eqnarray}
    {\cal D}(\tilde{A},\tilde{B}) &=& \Cdissip{A_{\rm e}}{B_{\rm e}}_x
    + \Cdissip{H_{\rm e}}{H_{\rm e}}^Q_x \cancor{\QPoiss{A}{Q}}{\QPoiss{B}{Q}}
    \nonumber \\
    &-& \Cdissip{A_{\rm e}}{H_{\rm e}}^Q_x \cancor{\QPoiss{H}{Q}}{\QPoiss{B}{Q}}
    \nonumber \\
    &-& \Cdissip{H_{\rm e}}{B_{\rm e}}^Q_x \cancor{\QPoiss{A}{Q}}{\QPoiss{H}{Q}}
    \nonumber \\
    &+& \Cdissip{A_{\rm e}}{B_{\rm e}}^Q_x \cancor{\QPoiss{H}{Q}}{\QPoiss{H}{Q}} .
\label{calDdef}
\end{eqnarray}
The standard classical dissipative bracket, $\Cdissip{\,}{\,}$, and the dissipative bracket for the coupling, $\Cdissip{\,}{\,}^Q$, are both symmetric and positive semidefinite, but they do not need to fulfill the usual GENERIC degeneracy expressing energy conservation; the conservation of energy is guaranteed by the form of Eq.~(\ref{calDdef}), which is actually motivated by energy conservation (that is why the last three compensation terms are needed). As the coupling between the quantum system and its classical environment is purely irreversible, the entropy of each subsystem is conserved by its own reversible dynamics, $\CPoiss{A_{\rm e}}{S_{\rm e}}=0$ and $\ave{\QPoiss{A}{S}}$ for arbitrary $\tilde{A} = (A, A_{\rm e})$.

The dissipative structure (\ref{calDdef}) can be generalized by summing up contributions from several different coupling operators $Q$, each of them coming with its own classical dissipative bracket $\Cdissip{\,}{\,}^Q$. The different irreversible processes could even be coupled.

From our fundamental equation (\ref{GENERIC}), we obtain the following equation for the evolution of averages in the quantum subsystem,
\begin{eqnarray}
    \frac{d\ave{A}}{dt} &=& \ave{\QPoiss{A}{H}}
    - \Cdissip{H_{\rm e}}{S_{\rm e}}^Q_x \cancor{\QPoiss{A}{Q}}{\QPoiss{H}{Q}}
    \nonumber\\
    && \qquad + \, \kB \, \Cdissip{H_{\rm e}}{H_{\rm e}}^Q_x
    \ave{\QPoiss{Q}{\QPoiss{Q}{A}}} .
\label{GENERICqma}
\end{eqnarray}
The classical environment is governed by the evolution equation
\begin{eqnarray}
    \frac{dA_{{\rm e},x}}{dt} &=& \CPoiss{A_{\rm e}}{H_{\rm e}}_x
    - \kB \, \Cdissip{A_{\rm e}}{H_{\rm e}}^Q_x \ave{\QPoiss{Q}{\QPoiss{Q}{H}}}
    \nonumber\\
    &+& \Cdissip{A_{\rm e}}{S_{\rm e}}_x + \, \Cdissip{A_{\rm e}}{S_{\rm e}}^Q_x
    \cancor{\QPoiss{H}{Q}}{\QPoiss{H}{Q}} . \qquad
\label{GENERICcla}
\end{eqnarray}
Note that the average of $Q$ in Eq.~(\ref{GENERICqma}) is not affected by the dissipation mediated by $Q$.

Equation (\ref{GENERICqma}) for averages of the quantum subsystem leads us to the following quantum master equation for the density matrix,
\begin{eqnarray}
    \frac{d\rho}{dt} = - \QPoiss{\rho}{H}
    &+& \Cdissip{H_{\rm e}}{S_{\rm e}}^Q_x \, \QPoiss{Q}{\QPoiss{Q}{H}_\rho}
    \nonumber\\
    &+& \kB \Cdissip{H_{\rm e}}{H_{\rm e}}^Q_x \, \QPoiss{Q}{\QPoiss{Q}{\rho}} .
    \qquad\qquad
\label{GENERICme}
\end{eqnarray}
The master equation (\ref{GENERICme}) is our fundamental equation for open quantum systems. The existence of a master equation implies that the evolution of all averages is formulated in a consistent way. In view of the definition (\ref{Atildef}), the second term in Eq.~(\ref{GENERICme}) will, in general, be nonlinear in $\rho$. This quantum nonlinearity caused by noncommutativity implies that our master equation cannot be of the usual Lindblad form (see, for example, Sec.~3.7 of \cite{BreuerPetru} for a discussion of nonlinear quantum master equations). The most natural linearization of the GENERIC master equation (\ref{GENERICme}) is obtained in terms of the symmetric anticommutator, $\QPoiss{Q}{H}_\rho \approx \Qantico{\QPoiss{Q}{H}}{\rho} /2$. Note, however, that linearizations spoil the thermodynamic structure and are hence not recommendable. The linearized equation then is of the Lindblad form, where a Lindblad operator with real and imaginary parts proportional to $Q$ and $\QPoiss{Q}{H}$, respectively, needs to be introduced and the Hamiltonian needs to be redefined.

Instead of Eqs.~(\ref{GENERICqma}) and (\ref{GENERICcla}) for averages, one usually prefers to solve the evolution equations for $\rho$ and $x$. The classical state variables $x$ may be considered as special cases of observables, and the quantum state variable $\rho$ can be parametrized by sufficiently many averages to obtain a closed set of equations. Note again that the counterpart of averaging quantum observables with the density matrix $\rho$ is the evaluation of classical observables in a certain state $x$.


\emph{Examples.}---As a simple example, we consider the motion of a particle of mass $m$ in a potential in one dimension. The position and momentum are given by $Q$ and $P$ with the canonical quantum Poisson bracket $\QPoiss{Q}{P}=1$, the potential is given by a function $V(Q)$. We use the position $Q$ as the coupling operator in Eq.~(\ref{GENERICqma}) because friction should affect the momentum $P$ of the particle, but not the position $Q$. The environment be a heat bath, the state of which be characterized by its energy $H_{\rm e}$. The thermodynamic properties of the bath are characterized by the thermodynamic relationship $S_{\rm e}(H_{\rm e})$, and we introduce the bath temperature by $1/T_{\rm e} = dS_{\rm e}/dH_{\rm e}$. The dissipative bracket is fully determined by a friction coefficient $\zeta$,
\begin{equation}\label{CalLegdb}
    \Cdissip{A_{\rm e}}{B_{\rm e}}^Q = \frac{dA_{\rm e}}{dH_{\rm e}}
    \, \zeta T_{\rm e} \, \frac{dB_{\rm e}}{dH_{\rm e}} .
\end{equation}
The equations for the first moments $\ave{Q}$, $\ave{P}$ and for the second moment $\ave{QQ}$ agree with those for the widely used Caldeira-Leggett high-temperature master equation (see Eqs.~(3.426)--(3.428) of \cite{BreuerPetru} for $\zeta = 2 \gamma m$ with a rate constant $\gamma$). In the equations for the second moments $\ave{PQ+QP}$ and $\ave{PP}$, however, there occur frictional contributions proportional to the canonical correlations $\cancor{P}{Q}$ and $\cancor{P}{P}$ rather than the averages $\ave{PQ+QP}/2$ and $\ave{PP}$ (see Eqs.~(3.429)--(3.430) of \cite{BreuerPetru}). The usual Caldeira-Leggett master equation \cite{CaldeiraLeggett83} (see, for example, Eqs.~(3.410) of \cite{BreuerPetru}) is recovered from the linearized GENERIC quantum master equation obtained after the previously mentioned approximation $\QPoiss{Q}{H}_\rho \approx \Qantico{\QPoiss{Q}{H}}{\rho} /2$ in the second term of Eq.~(\ref{GENERICme}).

As an example of a quantum optical master equation, we consider the decay of a two-level system described in terms of the Pauli matrices $\sigma_1$, $\sigma_2$, $\sigma_3$. We assume that this system is characterized by the Hamiltonian $H= (1/2) \hbar \omega \sigma_3$, where $\omega$ is the transition frequency of the free system, and exposed to black-body radiation of temperature  $T_{\rm e}$. The master equation for this system (see, for example, Eq.~(3.219) of \cite{BreuerPetru}) is recovered from the linearized GENERIC quantum master equation (\ref{GENERICme}) by choosing two coupling operators $Q$, namely $\sigma_1$ and $\sigma_2$, both with equal dissipation brackets $\Cdissip{H_{\rm e}}{H_{\rm e}}^Q_x = T_{\rm e} \hbar \gamma_0 / (4\omega)$, where $\gamma_0$ is the spontaneous emission rate. To obtain the quantum master equation for the decay of a two-level system in its familiar form, it is important to note $\QPoiss{\sigma_1}{H}= - \omega \sigma_2$ and $\QPoiss{\sigma_2}{H}= \omega \sigma_1$ and to introduce the rate $\gamma = 2 \gamma_0 \kB T_{\rm e} / (\hbar \omega)$.


\emph{Quantum perspective.}---In the above development, we have emphasized the thermodynamic perspective. We have identified the geometric structures required to formulate the thermodynamic evolution equations (\ref{GENERIC})--(\ref{calDdef}) for a quantum system coupled to a classical environment, which are the essence of this paper. With these equations at hand, one can concentrate on the formulation of detailed models in terms of plausible thermodynamic building blocks for specific real-world applications rather than on the theoretical derivation or justification of quantum master equations in simple special cases. Rather than heat baths only, one can treat all kinds of classical environments, including anisotropic and slowly changing environments. Particularly interesting is the special case where the environment evolves on a time scale comparable to the time scale on which it has a significant influence on the quantum subsystem.

Reaching the limit of perfectly isolated, purely reversible quantum systems may not only be experimentally difficult, but also from a conceptual point of view. The entropy operator in Eq.~(\ref{GENERIC}) is defined if, and only if, $p_n>0$ for all eigenstates of the density matrix. A possibly small but nonzero occupation of all quantum states is required for a meaningful thermodynamic description. Away from the idealization of a perfectly isolated system, the occurrence of probabilities in our theory is very natural. As long as the limit of perfect isolation cannot be reached, the interpretational problems of quantum mechanics seem to be absent.

We would like to point out briefly, that the theory of dissipative quantum systems in Eqs.~(\ref{GENERIC})--(\ref{calDdef}) should be considered as an ideal setting for addressing a number of fundamental problems. As renormalization procedures can be related to the elimination of degrees of freedom on short scales, it is only natural to expect that they are intimately related to statistical physics. In the context of self-similar polymers, it has actually been shown that dynamical renormalization can be performed in terms of a renormalization of the entropy and the dissipative bracket \cite{hco185}. The occurrence of a black hole entropy \cite{Bekenstein73,BardeenCarHawk73,Hawking74} becomes particularly interesting when it is coupled to other forms of entropy, including thermal entropy (see also \cite{hco180}). The dissipative quantum mechanics (\ref{GENERIC})--(\ref{calDdef}) would offer the proper framework for the description of such a coupling, too.

In short, this paper suggests that there is a deep give-and-take relationship between quantum mechanics and thermodynamics. In a number of respects, quantum mechanics with dissipation seems to be more natural than without dissipation.

Discussions with David Taj and J\'er\^ome Flakowski motivated me to move the topic of quantum dissipation high up on my priority list. I apologize for not exploring the pertinent literature more systematically but, as a beginner in this field, I preferred to be guided by the intrinsic beauty of the geometric approach.



\end{document}